# Performance of artificial neural networks in an inverse problem of laser beam diagnostics

Karol Pietrak[1*], Radosław Muszyński[2], Adam Marek[2] and Piotr Łapka[1]

[1]Institute of Heat Engineering, Warsaw University of Technology, 21/25 Nowowiejska St., 00-665 Warsaw, Poland

[2]Faculty of Power and Aeronautical Engineering, Warsaw University of Technology, 24 Nowowiejska St., 00-665 Warsaw, Poland

(*) corresponding author, e-mail: karol.pietrak@pw.edu.pl

## Abstract

Results are presented for the numerical verification of a method devised to identify an unknown spatio-temporal distribution of heat flux that occurs at the surface of thin aluminum plate, as a result of pulsed, high-power laser beam excitation. The presented identification of boundary heat flux function is a part of newly-proposed laser beam profiling method and utilizes artificial neural networks trained on temperature distributions generated with the ANSYS Fluent solver. The paper focuses on the selection of the most effective neural network hyperparameters (Keras, Tensorflow) and compares the results of neural network identification with Levenberg-Marquardt method used earlier and discussed in our previous articles.

## 1.    Introduction

High-power lasers are used in many areas of science and technology - for example in materials processing [1], medicine [2], defense [3] and materials characterization [4], [5]. One of their key characteristics considered in practical applications is the spatial distribution of energy density across the beam, also known as the beam profile [6]. The beam profile may be often described by the super-Gaussian function [7]. Three basic shapes of that function are recognized i.e. Gaussian, super-Gaussian and flattop (see Fig. 1). In laser-based devices it is frequently the case that light travels through various optical components before it leaves the device and interacts with its target. The profile of the working laser beam at the end of the optical track may differ from the profile at the source due to distortions caused by interactions of light with elements of the optical track such as mirrors, lenses and polarizers. The main cause of distortion is mechanical deformation of these elements which results from heating [8].

Verification of beam parameters and monitoring of their stability is required to assure controllable and repeatable operation of laser-based devices. It is typically achieved with the use of laser-beam profiling techniques [6], [8], among which destructive and non-destructive techniques may be distinguished. In destructive measurements, laser beams interact with solid targets (e.g. thermal paper, photographic films, acrylic blocks) and leave permanent changes which allow to identify some beam parameters [6]. Nevertheless, this type of setup is usually unsuitable for continuous measurements, and nowadays destructive techniques have mainly historical importance. Automated non-destructive techniques are typically commercialized and make use of various methods of beam attenuation before its interaction with the sensor. High power beams are particularly problematic to measure. Focused constant width lasers of power in the 1 W range and pulsed lasers in the 1 J range can easily damage scanning apertures, and beams in the kW range can damage beam samplers. One of the ways to overcome the power obstacle is the non-interceptive profiling proposed by Guttman [8]. Nevertheless, instruments based on that and other known methods are usually very fragile and expensive. Fragility makes it difficult to use them in field conditions, and high cost limits their use in the industry. Additionally, not all of them work for pulsed beams [10].

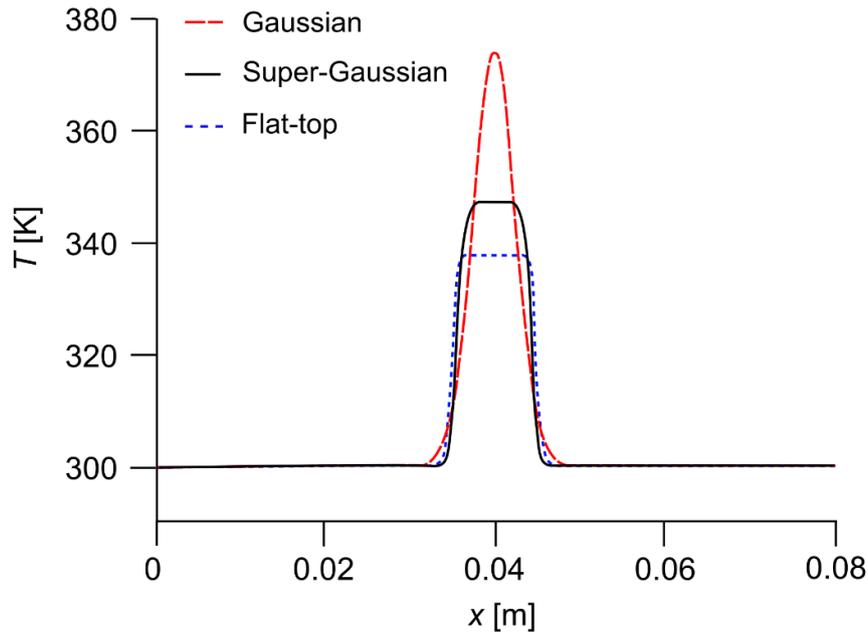

Fig. 1. Profiles of temperature at the surface of thin aluminum plate obtained by heating it with a high-power laser beams of three basic spatial energy distributions – Gaussian, super-Gaussian and Flat-top Adopted from [9].

In response to the need for less expensive and more robust testing methods applicable to pulsed laser beams, a new method has been recently proposed by Kujawińska et al. [9]. This method initially assumed utilization of fast infrared thermography and digital image correlation (DIC) [11] methods to capture temperature maps (thermography) and displacement maps (DIC) resulting from the interaction of the short laser pulse with a thin aluminum plate. The sensing of temperature and displacement is carried out for the rear surface of the plate – opposite to the heated surface (See Fig. 2). Kujawińska and her collaborators [12] assumed that their experimental setup may be used to identify four parameters of the laser pulse i.e. its spatial profile coefficient, beginning time, end time and power. In their technique, the density of heat flux at the sample surface, resulting from the laser strike, was assumed in the super-Gaussian form [7] and the impulse itself was treated as rectangular with respect to time.

The problem that arose may be classified as the inverse heat transfer problem (IHTP) of parameter estimation [13]. Pietrak et al. [12] and Łapka et al. [14] confirmed that useful results may be obtained with the aforementioned method even if only the thermal part of the problem is considered and the mechanical part (the measurement and simulation of displacements) is neglected. They tested the method based solely on temperature measurement using both simulated (artificial) [12] and real (physical) [14] measurement data. They obtained discrepancies between estimated and accurate parameter values of the order of 3.39% for artificial and 20-25% for physical experiments. The advantage of their technique is simple experimental setup employing a fast-infrared camera as the only required sensor. That fact makes the method less expensive and more resistant to failure than sophisticated commercial instruments. To solve the parameter estimation problem, Pietrak et al. and Łapka et al. used the Levenberg-Marquardt inverse algorithm with sensitivity matrices obtained by numerical simulation of the forward thermal problem. That approach has one drawback i.e. the presence of a significant delay between the physical measurement and the end of the data analysis. The delay is caused by the requirement to solve the forward problem many times using the Finite Volume Method (FVM) which may be time-consuming. To assure fast operation of the data analyzer program in field conditions, numerical simulation process may be moved to the time before the field measurements. This can be achieved with the artificial neural network (ANN) approach investigated in the current article.

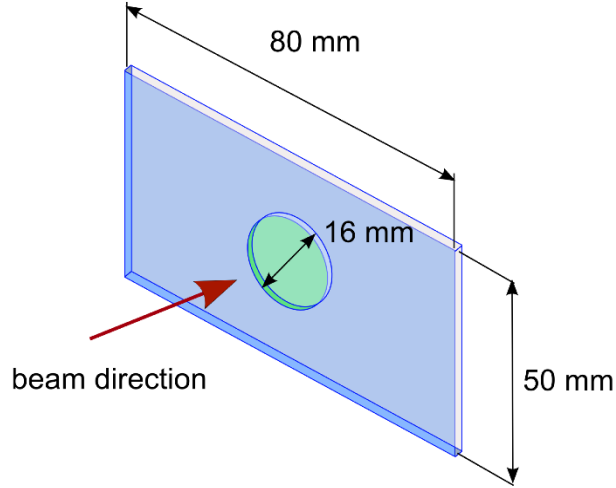

Fig. 2. The geometry of the laser-heated plate. The laser beam travels perpendicularly to the plate and hits its front surface which contains a circular hollow. Temperature is sensed at the opposite (rear) surface (no hollow).

The present approach to identification of spatio-temporal characteristics of laser pulse is based on the experimental setup proposed by Kujawińska et al. [9]. Similarly to the version considered by Pietrak et al. [12] and Łapka et al. [14] it includes the thermal problem and neglects the mechanical one. In other words, the unknown characteristics of the laser pulse are estimated only based on temperatures and not based on displacements. In the presented analysis, the aluminum sample was assumed to be heated by a short (0.2 – 2 ms) laser pulse traveling perpendicularly to the illuminated surface. The resulting heat transfer problem for the aluminum sample was modeled in the Ansys Fluent software. The spatial distribution of the boundary heat flux resulting from the laser heating was assumed in the super-Gaussian form [7] with rectangular temporal profile. It was assumed that the laser beam hits the front surface of the sample and temperatures are sensed at its rear surface. Temperature values at the rear surface were treated as inputs to the ANNs. The ANNs were trained using synthetic data from the Ansys Fluent model in which three parameters of the laser beam were varied i.e. the shape coefficient $p$, laser power $Q$ and pulse duration $t_{end}$ (the start time of the pulse was considered known). Training of the neural network using synthetic data allowed to avoid costly physical experiments and to estimate the potential of the proposed artificial intelligence solution for further development involving real measurement data. The ANN-based data analyzer may run on portable computers with limited resources and the results of laser beam parameters estimation may be obtained with minimal delay which makes the method advantageous.

Presented problem belongs to the class of inverse thermal problems of parameter estimation. Classical solution approaches to such problems were briefly reviewed by Beck and Woodbury [15]. The overview of inverse methods for determination of surface heat fluxes may be found in book chapters by Taler and Taler [16], [17]. Methods applied to problems of laser-solid matter interactions were discussed by Pietrak et al. [12]. ANNs, which constitute an important element of the current solution strategy, have been previously considered an interesting alternative method to solve general IHTPs alongside with genetic algorithms (GA), particle swarm optimization (PSO) and proper orthogonal decomposition (POD) [18]. Their successful application have been demonstrated in case of inverse and optimization problems involving conduction [19], [20], [21], [22], [23], [24], [25], [26], [27], radiation [28], [29], [30], [31] and convection [32], [33], [34], but also in power engineering [35], [36], [37]. In particular, Krejsa et al. assessed their potential and discussed various strategies regarding their application for inverse problem of heat conduction [19]. Detailed description of the theory of neural networks, oriented towards engineering and scientific applications, can be found in Samarasinghe's book [38].

## 2. Problem statement

### 2.1 Direct problem

In the direct problem considered in this paper, an aluminum (alloy *AW2017A T4*), rectangle-shaped sample with circular hollow in the center is struck by a short high-power laser pulse arriving perpendicularly to the plate from the side of the hollow (see Fig. 2). The reduction of sample thickness in the heating area decreases its temperature-damping quality and improves the accuracy of the subsequent parameter estimation as temperature sensing is carried out at the back side of the plate (opposite to the illuminated one) [14]. Front-face sensing of temperatures was also considered but it was finally rejected because of the risk of damaging the IR camera by the reflected laser light. The geometrical and thermophysical properties of the modeled body subjected to laser heating are grouped in Tab. 1.

Tab. 1. Geometrical and thermophysical properties of the heated sample

| Property | | Aluminum alloy AW2017A T4 | | | | | | |
|---|---|---|---|---|---|---|---|---|
| Temperature | [°C] | 30 | 50 | 100 | 150 | 200 | 250 | 300 |
| Width | [mm] | 80 | | | | | | |
| Height | [mm] | 50 | | | | | | |
| Thickness | [mm] | 1 | | | | | | |
| Milling (radius x depth) | [mm] | 8 x 0.5 | | | | | | |
| Density | [kg/m³] | 2700 | | | | | | |
| Thermal conductivity | [W/m/K] | 122.6 | 125 | 133.7 | 151.9 | 154.9 | 130 | 159.6 |
| Specific heat | [kJ/kg/K] | 0.902 | 0.911 | 0.943 | 1.046 | 1.039 | 0.832 | 0.995 |
| Surface emissivity | [-] | 0.3 | | | | | | |

The aluminum sample was considered homogeneous and isotropic. It was assumed that the measurement method should be non-destructive i.e. the temperature within the heated body should not exceed the melting point of aluminum (736 – 944 K depending on the alloy [39]). In such conditions the heat conduction process in the sample may described by the following equation:

$$\varrho c_p(T)\frac{\partial T}{\partial t} = div[\lambda(T)grad(T)], \tag{1}$$

where $\varrho$ denotes density, $c_p$ - specific heat, $\lambda$ - thermal conductivity, $T$ - temperature, $div[.]$ - divergence and $grad(.)$ – gradient operator. The boundary heat flux at the front surface (irradiated by laser beam) was assumed in the super-Gaussian form [7] as follows:

$$q_{laser} = \begin{cases} \varepsilon Q \dfrac{p2^{\frac{2}{p}}}{2\pi R_0^2 \Gamma\left(\frac{2}{p}\right)} \exp\left[-2\left(\dfrac{R}{R_0}\right)^p\right] & \text{for } 0 \leq t \leq t_{end} \\ \qquad\qquad 0.0 & \text{for } t > t_{end} \end{cases} \tag{2}$$

where $\varepsilon$ denotes surface emissivity (assumed $\varepsilon$ = 0.055 after matching the numerical model to experiments as presented in [14]), $Q$ - laser power, $p$ - curve shape coefficient, $R_0$ (assumed equal to 5 mm) - length scale over which the profile decreases to e⁻² of its axial value (in case of the Gaussian beam i.e. $p$ = 2), $R$ - distance from the center of the specimen and $\Gamma(.)$ – the Euler Gamma function. Let us note that by changing the parameter $p$ in equation (2) it is possible to obtain all beam distributions shown in Fig. 1, i.e. Gaussian ($p$ ~ 2) as well as Super-Gaussian ($p$ ~ 2-10) and Flat-top ($p$ ~ 10-100 and above).

On other walls boundary conditions were given by equation:

$$q_w = H(T_\infty - T_w) + \varepsilon\sigma(T_\infty^4 - T_w^4), \tag{3}$$

where subscripts $\infty$ and $w$ denote surroundings and the wall, respectively, $H$ – heat transfer coefficient at walls ($H$ = 5 W/m²/K was assumed based on the literature [40]) and $\sigma$ – Stefan-Boltzmann constant ($\sigma$ = 5.67 · 10⁻⁸ W/m²/K⁴). Initial temperature $T_0$ = 303 K was assumed within the whole body.

Thermal problem given by equations 1-3 have been solved using the FVM method in the commercial software *ANSYS Fluent 19.2*. Firstly, a 3D model of the problem was prepared, described in detail in [12] and [14], and its thermal parameters, such as surface reflectivity were finetuned based on physical experiments with laser and actual samples (the match of experimental and modeled temperature curves can be found in [14]). The geometry of the problem has been then transformed to 2D axisymmetric to reduce the computational cost. In the reduced model, the sample was treated as cylinder of radius 40 mm and height 1 mm with cylindrical hollow of radius 8 mm and depth 0.5 mm. The 2D model was meshed using regular grid containing 48450 rectangular elements with maximal aspect ratio of 271.46 and maximal skewness equal to $7.4 \cdot 10^{-4}$. The influence of grid sizing on result was checked and the analysis confirmed grid independence. The detail view of the computational mesh is shown in Fig. 3. The grid density was greater in the hollow section of the plate model (where laser excitation occurred) and lower in the remaining part. Finer grid was also placed near walls where heat diffusion starts and where temperature measurement points were located.

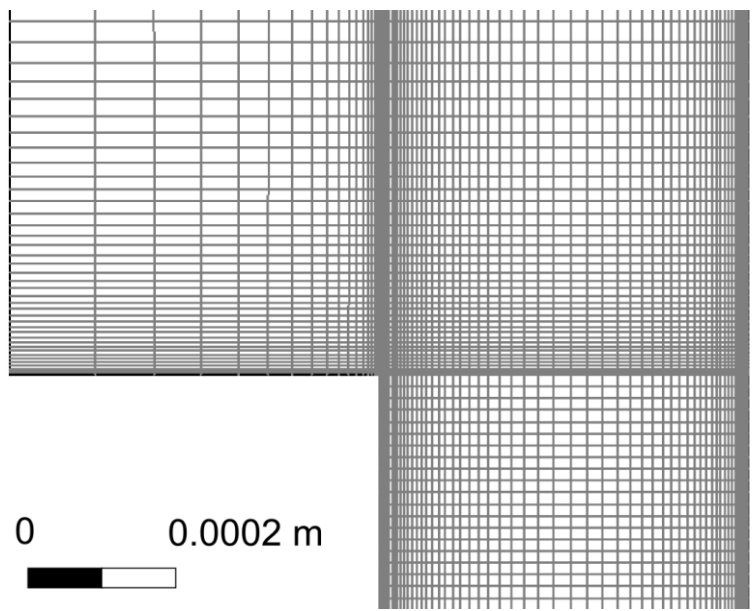

Fig. 3. Detail view of the computational grid

### 2.2 Inverse problem

#### 2.2.1 Estimated parameters (ANN outputs)

For the sake of the inverse analysis, it was assumed that the results of temperature measurements carried out at the rear side of the heated plate are available. The parameters estimated by the network were:
1) laser power $Q$,
2) spatial profile coefficient of the beam $p$ and
3) pulse end time $t_{end}$

The start time was assumed known: $t_{start} = 0$.

#### 2.2.2 Observed parameters (ANN inputs)

Before applying the ANNs it was necessary to select the inputs. Temperature measurements in real experiments are performed by fast IR cameras which leads to availability of full temperature maps of the investigated area, acquired with temporal step of the order of 0.27 ms and spatial resolution of 80×64 pixels [41]. The large array of data resulting from the IR measurement must be limited to become an input array to ANN. Appropriate selections were made based on earlier analyses of sensitivity [12]. It was assumed that the laser pulse end time $t_{end}$ belongs to the range [0.2, 2] ms. The duration of temperature measurement was limited to 16 ms, which assured that the decrease of temperatures due to heat loss to the surroundings is captured. The area of temperature sensing was limited to the central region of the wall placed vis-à-vis the heated area, where problem-significant temperature changes were occurring.

The input vector of measurements used to train the ANNs was constructed from two subsets:

1) temperature change history at the central point of the sample (see Fig. 4);
2) temperature values along the sample radius for a single time instant $t = 12$ ms.

The first subset of the vector reflected the temporal profile of the thermal response of the sample and the second one – its spatial profile. Such vector was sampled with 3 different resolutions (n = 30, 60, 134 points) as the influence of the sampling resolution on the retrieval error was examined. Spatial and temporal responses were sampled equally e.g. for 60 points 30 represented the temperature profile along the radius and the other 30 – the change of temperature in the central point.

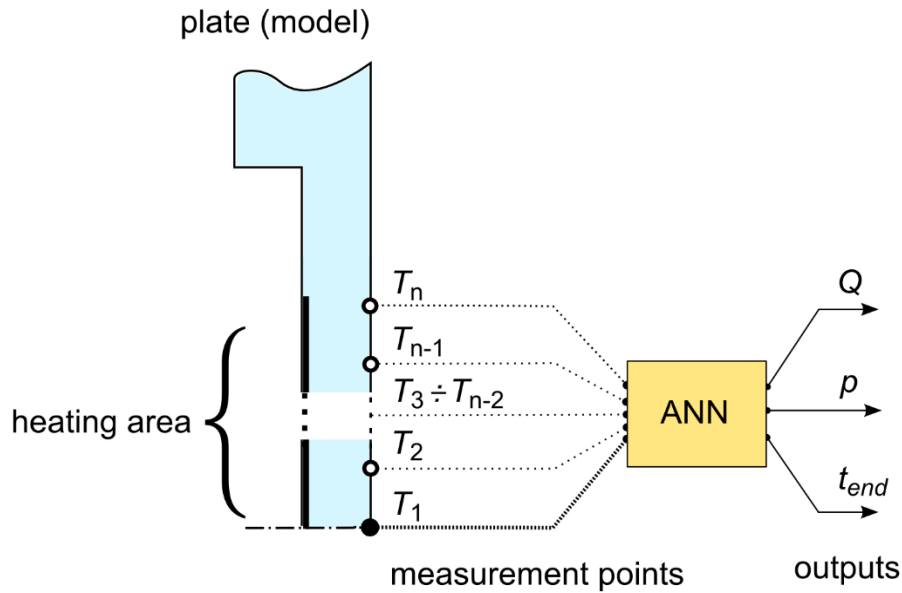

Fig. 4. The schematic of considered axisymmetric plate model with locations of measurement points from which the inputs to artificial neural networks were taken. The ANN and estimated pulsed laser beam parameters (outputs) are also shown.

Raw data sets for ANNs were generated by simulating the axisymmetric FVM model with different values of input parameters $Q$, $p$ and $t_{end}$. The range of pulse durations and laser power was limited to assure that melting temperature is not reached. For both scenarios, values of parameter $Q$ (in kilowatts) were taken from the set

$$\{5, 6, 7, \ldots, 20\}, \tag{4}$$

values of parameter $p$ (dimensionless) from the set

$$\{2, 8, 14, \ldots, 98\}. \tag{5}$$

and values of parameter $t_{end}$ (in ms), from the set

$$\{0.2, 0.4, 0.6, \ldots, 2\}. \tag{6}$$

The whole database contained 2720 cases.

## 3. Methods

### 3.1 Neural computation

#### 3.1.1 Training methods

The neural networks were programmed in Python 3.7 using open-source libraries Keras and Tensorflow [42]. Initially, weights were set to random numbers. In the whole study the maximal number of hidden layers was limited to two. The procedure of searching for the best ANN configuration was divided into two stages – initial and final. Firstly, the influence of network topology was tested with other hyperparameters constrained to arbitrarily-chosen ones (see section 4.1). Then, two single- and double-hidden layer topologies which achieved the best performance (i.e. 60-120-3 and 60-120-240-3, respectively) were used combination with the same arbitrarily-chosen hyperparameters treated as fixed, while different optimizers, activation functions and data preprocessing methods were tested one at a time, successively (sections 4.2 – 4.4). This initial stage allowed to draw conclusions for the second stage of tests in which allowed to narrow down the options and select the recommended configuration of hyperparametrs for the given application.

Among the training algorithms tested were all gradient descent optimization algorithms available in the Keras library, i.e. Stochastic gradient descent (SGD), Root Mean Square Propagation (RMSProp), Adagrad, Adadelta, Adam, Adamax and Nadam. Aforementioned algorithms, among others, were thoroughly explained by Ruder [43]. Tested activation functions included Sigmoid [44], [45], Tanh [44], [45], Relu [44], [45], Elu [45], [46], Selu [45], [47], Softmax [45], Softplus [45], Softsign [45] and Hard_sigmoid [45].

To avoid overfitting, the 'EarlyStopping' function [48] was used which stops training at an optimal point, when a neural network does not learn anymore. The function monitors validation loss according to the selected metric, which in this experiment was the MSE (mean squared error) [49]. The only customizable parameter was 'patience' which tells how many epochs must be executed without improvement to break the process of training. This is because usually learning is not linear and after some epochs of decreasing performance, there might come epochs with even better results. For all trainings of neural networks mentioned in this paper, the parameter of 'patience' was set to 100 epochs. This is because according to our observations, insufficient number of epochs was often leading to small accuracy while too big number brought no improvement.

#### 3.1.2 Topology investigation

Although the concept of machine learning is not new, there has not been found a method to calculate analytically the best number of hidden layers or number of nodes for a model [50]. Every problem is specific, so only suggestions can be found in literature but not the final answer. Testing many possibilities is a problematic solution because of extremely large number of combinations and long time of training a single model. In general, the more nodes and hidden layers are used, the more complex functions may be approximated. On the other hand, large networks take longer to train and are more susceptible to overfitting [51].

The Hecht-Nielsen theorem states that a network with only one hidden layer and $2i + 1$ nodes (where $i$ denotes the number of inputs) is enough to approximate any continuous function, provided that special activation functions are used [52]. Kurkova suggested that a second layer can be added to obtain equal capabilities with classical activation functions [53]. Later, Cybenko shown that an ANN with one hidden layer and the commonly used sigmoid activation function can approximate any continuous function of $n$ variables real variables with support in the unit hypercube [54]. Given these rules, it was reasonable to start experimenting with a single hidden layer and check how adding more layers affected accuracy and generalization ability.

To gain an initial insight regarding an optimal topology for the current problem, architectures with one and two single layers were tested, starting with one hidden layer of 10 nodes, increasing gradually to 1200 nodes, and then expanding to two layers, 10 neurons each, increasing both to nearly 1000 nodes (e.g. 800 in the first and 1000 in the second). In these experiments, the optimization algorithm (Adam [43]), stopping criteria and activation functions (sigmoid for hidden layer, non for output layer) were the same to obtain a useful comparison. The inputs were noiseless, and their number was always 60. In later experiments selected topologies with one and two hidden layers were tried with different optimizers, activation functions and data preparation methods.

### 3.1.3 Data preparation

Due to the characteristics of algorithms applied to update weights and biases, it is generally observed that input data preprocessing (rescaling, normalization and standardization [55]) is beneficial for the neural network performance [55], [56], [57]. Preprocessed data leads to faster training and better accuracy of estimations. In the present case, the ANN models did not work with raw data. Therefore, various preconditioning methods for the input temperature data were tested, i.e. min-max standardization to <-1,1> [58], dividing by the median [58], subtracting the median [59], and subtracting the midrange [60].

Furtherly, also the expected values (in our case laser pulse parameters) may be transformed to improve the prediction accuracy in machine learning [57]. It was decided to compare the accuracy without modification of the expected values (raw data) and with outputs scaled to the similar range as the inputs.

## 3.2 Performance assessment methods

### 3.2.1 Verification methods

For the experiments, the data was randomly divided into 3 groups:

    2000 samples – training data,
    420 samples – validation data,
    300 samples – test data.

The first subset was used for training. The second one was monitored during training to detect overfitting. The last set was not used for training and was used to assess the network performance when it was presented with unknown data.

In heat transfer problems, the measured data is always noisy. Such cases were considered in this work to model the reality more realistically. In order to check how noise affects the performance, the data sets were modified by addition of random noise of uniform distribution. It was added with four levels of amplitude: 0.1 K; 0.25 K; 0.5 K and 1K. The formula (7) presents the basic modification of a sample in the dataset, where: $i, j$ – index numbers of feature and sample, $T_s^{i,j}$ - the temperature simulated in ANSYS fluent software, $T_f^{i,j}$ - the temperature used as final feature, $h$ – noise randomly generated from fixed range of given amplitude.

$$T_f^{i,j} = T_s^{i,j} + h. \tag{7}$$

Unfortunately, a neural network trained on a dataset with given level of noise does not provide highest performance on the sets with different levels of noise. Due to that, models were retrained to accommodate for data with different levels of noise.

### 3.2.2 Error metric

The level of agreement between the test datasets and network predictions was calculated using Mean Absolute Percentage Error (MAPE) which can be described by the formula (8), where: $E_i$ – the value estimated by ANN, $C_i$ – the correct value, z – the number of estimations, $|.|$ - the absolute value and $i \epsilon$ 1,2,...,z – the index of estimation. It gives the information about the mean deviation of estimations from accurate parameter values expressed as percentage.

$$MAPE = \frac{1}{z}\sum_{i=1}^{z}\frac{|E_i - C_i|}{|C_i|} * 100\% \tag{8}$$

The error was measured for all 3 laser parameters separately but because there were no significant differences observed between them, for every case only the average MAPE was considered.

## 4. Results of initial studies

### 4.1 The initial topology investigation

In the first stage of topology optimization, multiple single- and double-layer topologies were verified systematically with the number of inputs constrained to 60, Adam optimizer, sigmoid activation function for hidden layers, and non-sigmoid for the output layer. Fig. 5 shows the behavior of the estimation error as the number of neurons in the single hidden layer case is increased. It may be seen that the MAPE decreases steadily until the number of neurons in the hidden layer is equal to the number of inputs (60). The local minimum of error (MAPE = 3.82%) occurred for 120 neurons. It had been concluded that further increase of the layer size was not effective as the error fluctuated above 120 neurons with no clear improvement. The global minimum occurred for 700 neurons (MAPE = 3.33%), but it was be regarded as coincidence due to the fluctuating nature of the error value in that range.

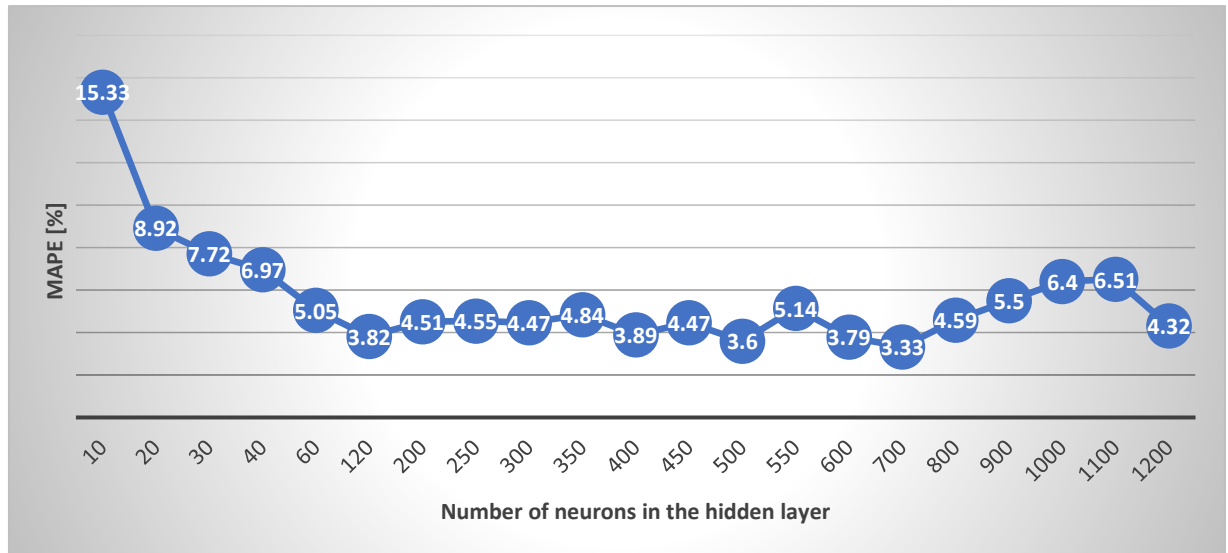

Fig. 5. Relation between estimation error and number of neurons in the hidden layer in case of one hidden layer

Surprisingly, the results for two hidden layers, with the same settings, were generally worse. The sizes of the first layer in tested configurations were the same as for the single-layer cases and the sizes of the second layer were selected to be of the similar order (i.e. 60-120, 1000-800 etc.). A total number of 24 topologies with two hidden layers was examined, and no clear minima could be identified. However, the results were best if both layers contained 60 neurons and more. The overall estimation accuracy was lower than for a single hidden layer. The median and mean of error for single hidden layer experiments were 4.59 and 5.56% respectively, whereas for the cases with two layers these were 5.78 and 6.65%.

*4.2 Optimizers investigation*

At the next stage of research, all the optimizers were tested for both single-hidden layer and double-hidden layer topologies which achieved the best results (i.e. 60-120-3 and 60-120-240-3, respectively). Three simulations were performed for each optimizer to verify the repeatability. Sigmoid activation functions were used for hidden layers, and non-sigmoid for output layers.

In case of one hidden layer, the best repeatable results (MAPE around 5%) were obtained with Adam and Adamax optimizers, as can be seen in Fig. 6. When the optimizers were tested with the double-hidden layer network, even better results were obtained (Fig. 7). In those experiments, the best accuracy was achieved with the Adagrad algorithm (MAPE around 2.5%). The typical number of epochs to complete the training for that algorithm was generally higher than for other algorithms, both in case of single (5000-50000 epochs for Adagrad vs 500-2000 for other optimizers) and double hidden layer (2800-5000 epochs for Adagrad vs 250-900 for other optimizers).

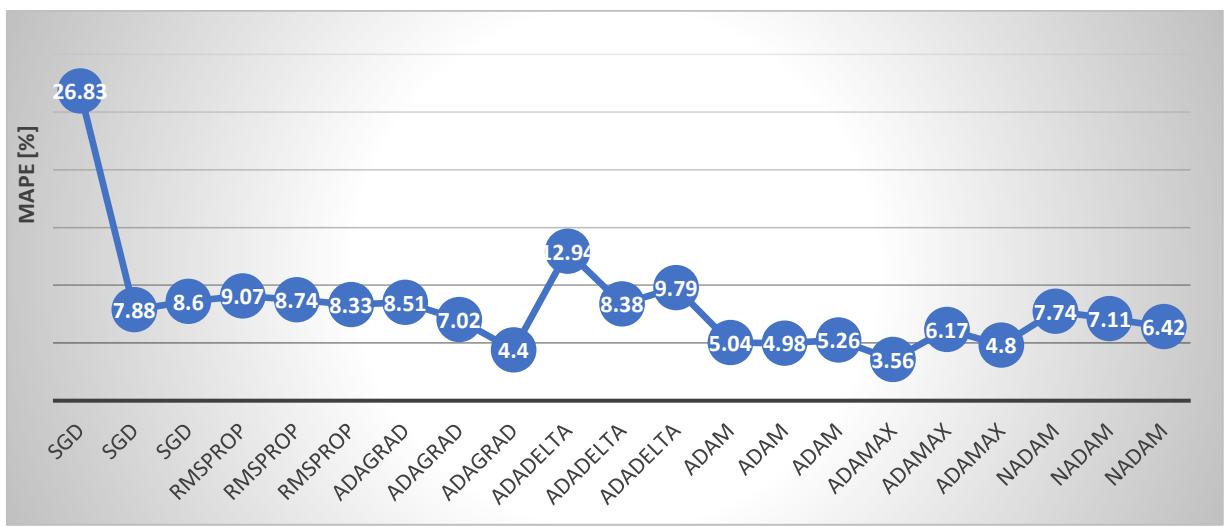

Fig. 6. MAPE error for different optimizers tested on ANN with single hidden layer

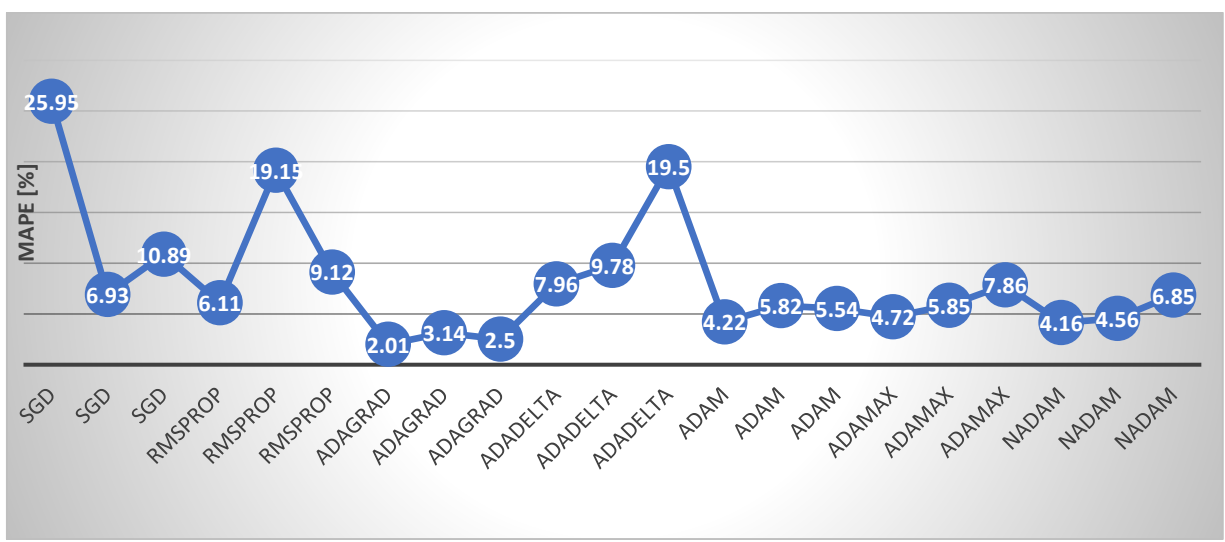

Fig. 7. MAPE error for different optimizers tested on ANN with two hidden layers

## 4.3 Investigation of activation functions

The choice of activation functions may influence the ANN's approximation ability and accuracy [45], [52] and due to that it was decided to test all activation functions available in the Keras library. The functions were compared on networks with topologies which provided the best results in the initial topology study i.e. 60-120-3 and 60-120-240-3. The comparisons of estimation accuracy are shown in Fig. 8 and Fig. 9. The MAPE errors shown are averaged from 3 experiments. The results for the Softmax function are not shown, because the error in that case was extremely high (90 and 136% respectively for single and double hidden layer). Remaining activation functions gave results of comparable order, with Softplus function being the best in case of one hidden layer and Relu - in case of two hidden layers.

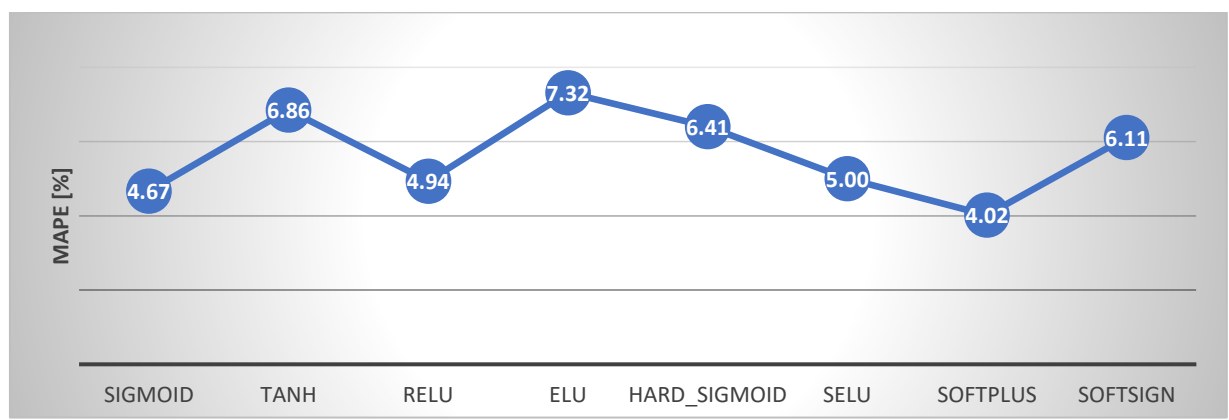

Fig. 8. Estimation errors for ANN with one hidden layer and different activation functions. The results are averaged from 3 experiments.

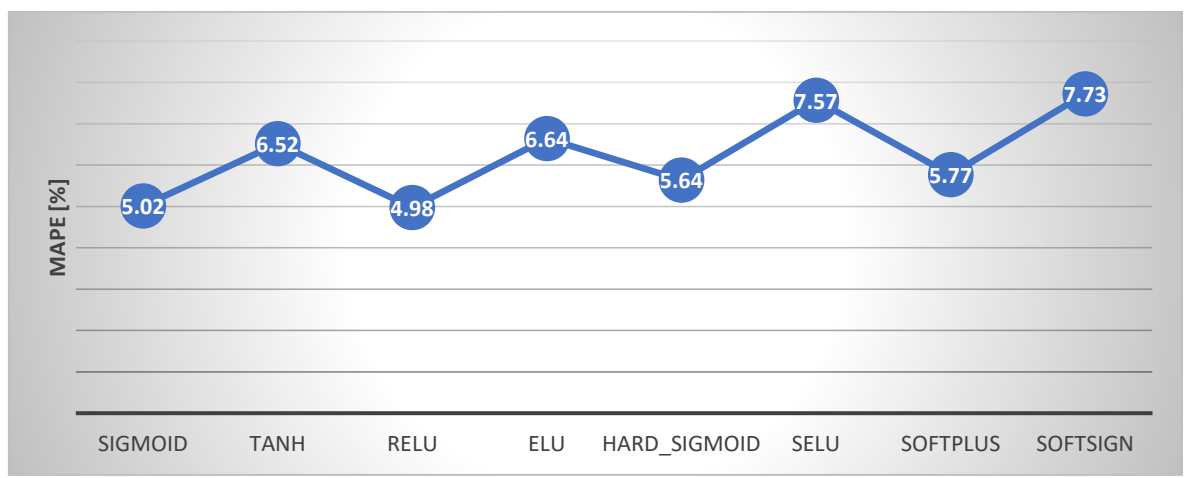

Fig. 9. Estimation errors for ANN with two hidden layers and different activation functions. The results are averaged from 3 experiments

## 4.4 Investigation of data preprocessing methods

In order to find the best preprocessing methods, few series of experiments were conducted. In the first series, input data transformation methods (i.e. min-max standardization, subtraction of the median, subtraction of the midrange, and division by the median) were compared. Three simulations were performed per every method to obtain averaged results shown in Fig. 10. Parameters of neural networks in the experiments were: topology 60-120-3, Activation functions -sigmoid for the hidden layer, non-sigmoid for the output layer, optimizer- adam.

It can be seen in Fig. 9, that simply dividing the measured temperatures (in kelvins) by their median was not effective (estimation error 68 to 98%), while other methods, which shifted the center of the data to zero, gave considerably better results. Among those methods, subtracting the median resulted in the smallest error (around 5.4%).

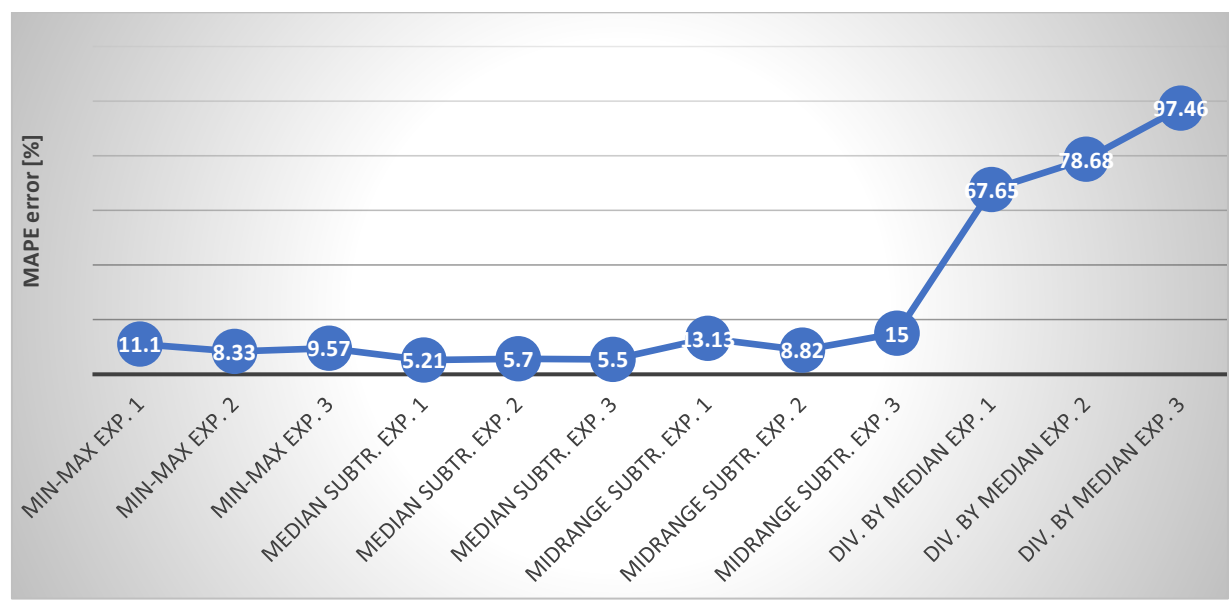

Fig. 10. Comparison of estimation errors for various methods of input data transformation. For each method, three experiments were performed. Network topology: 60-120-3

Secondly, the outputs had been scaled to possess similar range to the inputs, i.e., laser power in W was divided by 10 and shape coefficient $p$ was divided by 20, whereas the pulse duration remained unchanged. The estimation accuracy in such case was compared with the case of unaltered outputs (raw data). The effect was checked on five experiments for each type of data (altered/unaltered). As can be seen in Fig. 10, scaling of the output data resulted in profound increase of accuracy. Fig. 11 shows results for a network with two hidden layers (60-120-240-3). The single-hidden layer network (60-120-240-3) gave similar results.

In one experiment the pulse duration parameter (in ms) was multiplied by 10 which resulted in all output parameters having similar magnitudes (1-100). However, it was still much greater than the magnitude of inputs. The study shown that this type of treatment is ineffective. The range of input and output data should be similar. Subtracting of the median from the scaled output data was also tried in order to shift the center of the data to zero, but it resulted in loss of stability (MAPE for one parameter begin tending to infinity).

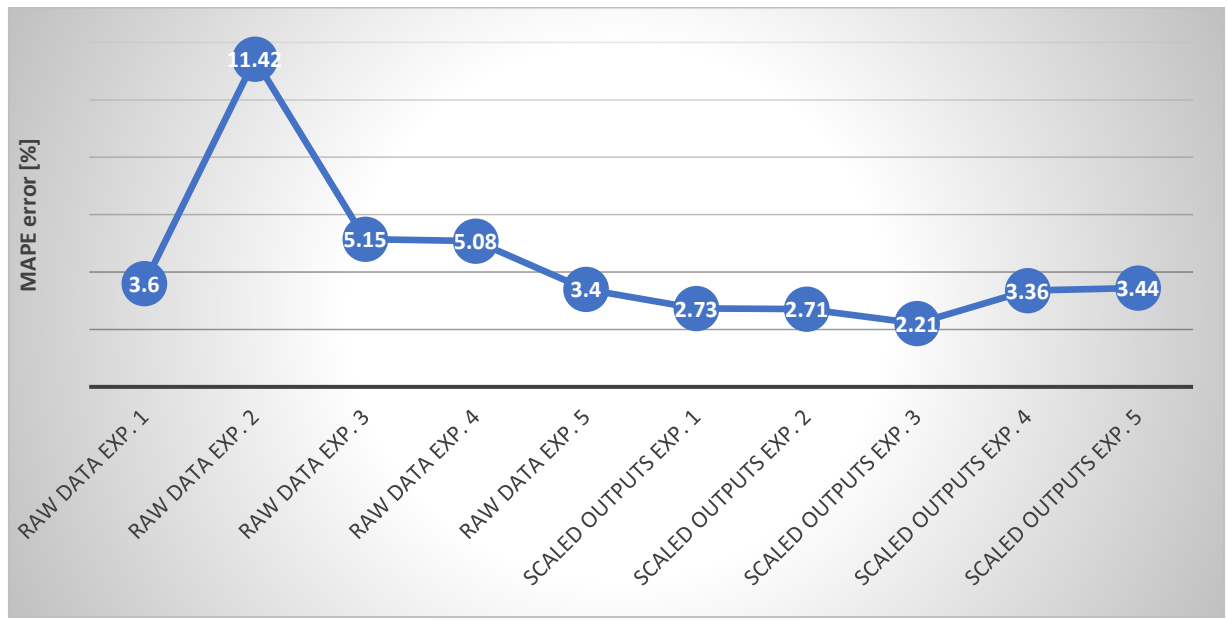

Fig. 11. Comparison of estimation errors for raw and scaled inputs for ANN of topology 60-120-240-3.

## 5. Results of the final study

### 5.1 Selection of the best ANN configuration

Although the initial topology study shown that a single-hidden layer network might be the best, changing the optimizer to Adagrad caused the error to drop significantly for a network with two hidden layers (see Fig. 7). Relu and sigmoid activation functions were selected for additional tests because they demonstrated the best performance in double-hidden layer networks, and good performance in single-hidden layer networks (Figs 8 and 9).

Additional experiments have shown that the best accuracy is assured if both input and output layer operate with relu activation function (configuration relu-relu resulted in 2% error, as opposed to 2.25% for sigmoid-relu and 3.4% for sigmoid-sigmoid). It was hard to point out the optimal topology, apart from the fact that the number of neurons in the hidden layers should be greater than, or equal to, the number of inputs. Based on additional experiments with Adagrad optimizer and relu-relu activation functions, the architecture 60-1000-1000-3 was identified as best and selected for further studies. The topology with high number of neurons in hidden layer was expected to perform good in the upcoming simulations with number of inputs increased to 134.

### 5.2 The effect of noise and measurement resolution

The dataset of simulated temperatures from the Fluent software was characterized by high spatial and temporal resolution of the data. This feature allowed to increase the number of ANN inputs to 134 (increased measurement resolution). Both spatial and temporal resolutions were increased equally. The influence of increase and decrease of the number of inputs on ANN retrieval accuracy was tested using the best ANN configuration X-1000-1000-3 (see Fig. 12).

At this stage, the input data were modified by the addition of uniformly-distributed pseudorandom noise (eq. 7) of different amplitudes. The influence of temperature measurement resolution and the level of noise on the accuracy of estimations is shown in Fig. 11. To generate the graph, each configuration was trained three times and the arithmetic mean of obtained MAPEs was calculated. The scatter of results was very small (within $0.1 - 1\%$ range). It was observed that ANN estimation results are significantly more repeatable if the training data are noisy, compared to non-noisy data.

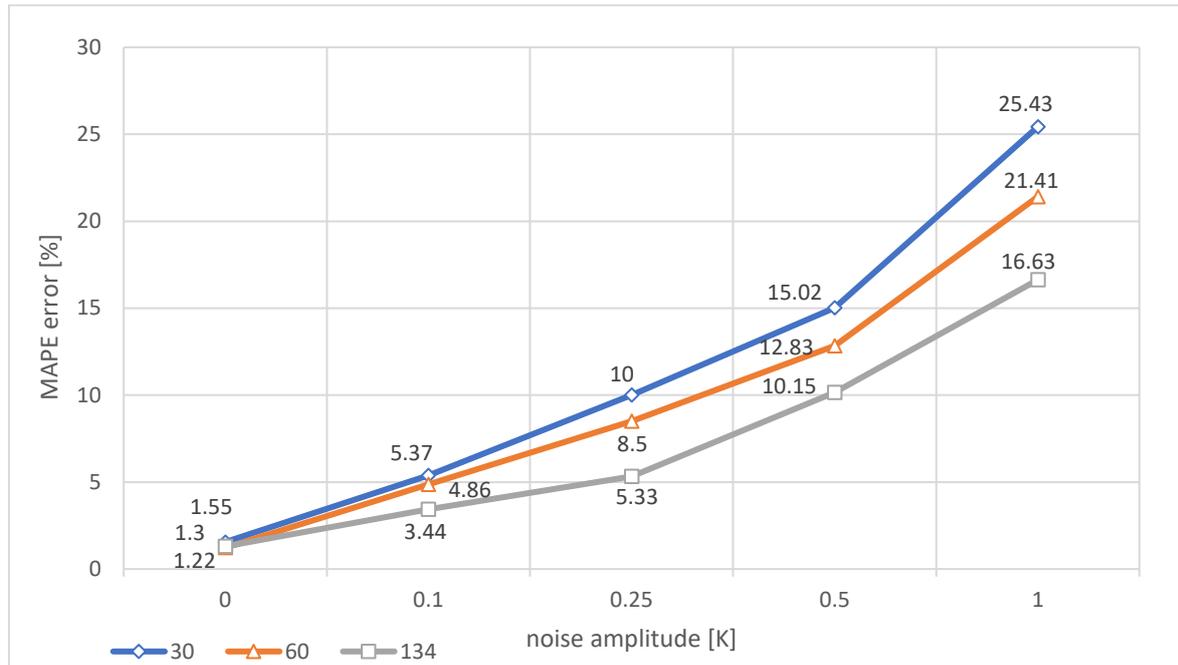

Fig. 12. Dependence of the parameter estimation error on the number of inputs and the level of noise in the measured temperatures for the best ANN configuration. Different lines represent different numbers of inputs

## 6. Discussion

The lowest MAPE error achieved during all tests was 1.22%, but it was for non-noisy data which represents a non-realistic case. For the noise amplitude of 0.1K the best result was MAPE error equal to 3.44%, for 0.25K it rose to 5.33%, for 0.5K it was 10.15% and for 1K 16.63%. The other important factor is the number of input points i.e. the number of temperatures measured on the plate during one test. For non-noisy data, the MAPE error in all 3 cases (30, 60, 120 input points) was nearing. However, the higher the noise level, the bigger the difference between them. The models with more inputs had evidently better performance for noisy datasets.

To judge the quality of the finally obtained result, it should be compared with similar research and analyzed from different perspectives. The same inverse problem was previously solved using gradient-based Levenberg-Marquardt (LM) technique in which the heat transfer model in Fluent was directly simulated to obtain the sensitivity matrices [12]. With that technique retrieval errors as low as 1.55% were obtained when the level of noise was 1K. For the same noise, the current method yielded much greater error (16.63%). Nevertheless, there is a substantial difference – the LM method uses direct simulation of the heat transfer model to take iterative steps and gradually minimize the difference between modeled data and experimental data in the full range of possible parameters, whereas the ANN method generalizes based on a limited number of cases generated for some fixed configurations of parameters. A neural network can give a rough estimate of the unknown parameters immediately even if the details of the heat transfer model are not known, assuming that it is trained with real experimental data. The LM method provided better result, but it required the knowledge of the heat transfer model and long computation times. The machine learning analyzer may be advantageous in situations where immediate estimate is needed or the model of the phenomenon is not fully known (it can rely solely on the measurement data). If time is not very important, the less accurate but immediate parameter estimation method can be used to obtain the initial estimate, before a more accurate estimate from the second method is available.

It must be noted that the accuracy of the neural network method can be increased by raising the signal-to-noise ratio. The signal to noise ratio is especially low if laser pulses of low power and low pulse duration are measured. In this case, the temperature increase sensed at the back surface of the plate is low. To increase the accuracy, thinner measuring plates may be used. In this way, specialized plates may be used for different ranges of laser pulse energy. Such design was not applied in current preliminary tests but should be considered in future designs.

Ideally, the measuring plate should be as thin as possible because it guarantees greater changes of temperature in the sensing area and will reduce the ill-conditioning of the inverse problem. Nevertheless, thinner plate requires less energy to start the melting process. Although the preliminary tests have shown that laser pulse parameters may be retrieved even in case of plate melting [61], a non-destructive method (without melting) is more desirable (due to material savings and easier maintenance of the instrument). In that case, material such as tungsten, withstanding higher temperature and possessing good thermal diffusivity, can be used to produce thin measuring plates, which should assure better signal-to-noise ratio.

Other methods to reduce the influence of noise are software noise cancellation or repeated measurements. In the latter method, the measurement of temperature response to a laser pulse is repeated and the data from those multiple measurements are averaged which reduces the noise (because the expected value of pure noise is typically zero). The noise reduction is greater for greater number of experiments used for averaging. According to our simulated experiments, averaging of data from 5 repeated experiments allows to half the MAPE error in case of the ANN estimation method (i.e. from 16 to 8%).

For the method to work best, a reduction of thermal imaging noise to minimum must be assured. In the physical experiments conducted to test the gradient-based method [14] the typical level of spatial noise in IR images was about ±0.5 K, and temporal noise ±0.25K, which corresponds to ANN estimation error of around 8%, based on Fig. 11. However, in these experiments no special attention was directed at assuring noise reduction. Shielding of the elements of the optical track and stabilization of the electric power sources should result in a further increase of the method accuracy. The levels of noise in the captured data [14] were pretty high for a modern Stirling-cooled IR camera, which suggests that better signal-to-noise ratio is still possible [57].

Now moving to comparisons with a broader spectrum of works, it should be said that the ANN approach was previously successfully applied in heat transfer engineering to solve various problems of estimation and prediction. For example, Czél et al. [23] performed numerical simulations regarding simultaneous estimation of temperature-dependent volumetric heat capacity and thermal conductivity in a solid material and obtained 1.216% average error for noisy inputs. Their ANN structure contained 34 output and 34 or 162 input neurons. Romero-Méndez et al. [33] reported 2.46% error in the problem of convective heat transfer coefficient estimation in evaporative mini-tubes based on the knowledge of saturation temperature, refrigerant mass flow rate and applied heat flux.

A case similar to the current one, involving a radiative source and temperature measurements, was presented by Mirsepahi et al. [29] who used shallow neural networks trained with LM to estimate the temporal change of irradiative heat generated by a heat lamp in a batch dryer based on the temperature history measurement on one of the surfaces of the enclosure, as measured by a single thermocouple. In their first paper, the mean of absolute error (MAE) was 43.00245 W for the input heat in the range of 0–2 kW. For an average input heat of 1 kW this translates to 4.3% MAPE. Data from physical experiments were used to train the networks, and the reported error of temperature measurement was ±1K. In the subsequent research [30] genetic algorithms were applied to optimize the number of neurons in hidden layers, which allowed to decrease the MAE to 35.05 W (MAPE ~3.5%). In the next paper [28], the tested physical system was extended to a dryer with two chambers, each heated by a single lamp. Each chamber was monitored with one thermocouple. The presented data allow to estimate the MAPE in that case to be around 3%.

At the same level of sensor error, the MAPE in the present solution is much greater (16%), which can be explained by the fact that the current problem suffers more from ill-conditioning. For the irradiative dryer, the heat transfer between the lamp and the thermocouple is confined within the enclosure, which makes it well shielded from ambient noise, and the sensor is separated from the source only by air. In the current solution, the heat source is obstructed by a solid body, which is necessary to prevent sensor damage. The radial spread of heat within the plate and its thermal inertia greatly weaken the effect measured at the opposite surface which results in ill-conditioning. It is still possible to overcome these problems and increase the accuracy in future designs by measuring the heat effects directly at the irradiated surface.

## 7. Conclusions

- The method is attractive as it can work purely on measurement data, without the heat transfer model (which parameters are often hard to establish). Once the ANN is trained, estimations are given immediately.
- The best results were obtained when the number of neurons in the hidden layers was more than or equal with the number of inputs.
- Changing the optimizer from Adam to Adagrad introduced a significant diminishing of error (from around 5 to 2.5% for undistorted data) and unlocked the potential of the second hidden layer.
- During the testing of different activation functions, it was found that the Softmax function leads to disproportionally large errors (90, 136%). Errors while applying other functions were far smaller.
- Both subtracting the median and scaling the inputs to similar range to outputs resulted in a profound increase of estimation accuracy.
- Increasing the number of available inputs is especially effective in lowering the estimation errors for noisy data.
- For the levels of noise measured in physical experiments (0.25-0.5 K), the accuracy of the ANN parameter estimation method is between 5 and 10%.
- The method accuracy may still be increased in future designs, either through utilization of plates of different thicknesses for different ranges of pulse energy or by changing the temperature measurement location to the irradiated front surface. Among these two solutions, the second one appears as one with greater potential.

## CRediT authorship statement

## Acknowledgements

The funding for this research was obtained from the statutory funds of the Faculty of Power and Aeronautical Engineering of Warsaw University of Technology.

## References

[1] A. K. Dubey and V. Yadava, "Laser beam machining—A review," *Int. J. Mach. Tools Manuf.*, vol. 48, no. 6, pp. 609–628, 2008, doi: 10.1016/j.ijmachtools.2007.10.017.

[2] Q. Peng *et al.*, "Lasers in medicine," *Reports Prog. Phys.*, vol. 71, no. 5, p. 56701, 2008, doi: 10.1088/0034-4885/71/5/056701.

[3]     G. P. Perram, M. A. Marciniak, and M. Goda, "High energy laser weapons: technology overview," *Proc. SPIE*, vol. 5414, pp. 1–25, 2004, doi: 10.1117/12.544529.

[4]     K. Pietrak and T. S. Wiśniewski, "Methods for experimental determination of solid-solid interfacial thermal resistance with application to composite materials," *J. Power Technol.*, vol. 94, no. 4, pp. 270–285, 2014.

[5]     L. Bruno, "Mechanical characterization of composite materials by optical techniques: A review," *Opt. Lasers Eng.*, vol. 104, pp. 192–203, 2018, doi: 10.1016/J.OPTLASENG.2017.06.016.

[6]     C. B. Roundy, "Current technologies of laser beam profile measurements," in *Laser Beam Shaping*, F. M. Dickey and S. C. Holswade, Eds., New York: CRC Press, 2000.

[7]     J. A. Hoffnagle and C. M. Jefferson, "Design and performance of a refractive optical system that converts a Gaussian to a flattop beam.," *Appl. Opt.*, vol. 39, no. 30, pp. 5488–5499, 2000, [Online]. Available: http://www.ncbi.nlm.nih.gov/pubmed/18354545

[8]     J. L. Guttman, "Noninterceptive Beam Profiling of High-Power Industrial Lasers," *Laser Tech. J.*, vol. 12, no. 5, pp. 20–23, 2015, doi: 10.1002/latj.201500035.

[9]     M. Kujawińska, P. Łapka, M. Malesa, K. Malowany, M. Prasek, and J. Marczak, "Investigations of high power laser beam interaction with material by means of hybrid FVM-FEM and digital image correlation methods," in *Proc. SPIE 10159*, 2016, p. 1015916. doi: 10.1117/12.2265745.

[10]    O. Aharon, "High power beam analysis," *Proc. SPIE 8963, High-Power Laser Mater. Process. Lasers, Beam Deliv. Diagnostics, Appl. III*, vol. 28, no. 5, p. 89630M, 2013, doi: 10.1117/12.2036550.

[11]    M. A. Sutton, J. J. Orteu, and H. Schreier, *Image Correlation for Shape, Motion and Deformation Measurements*. Springer US, 2009. doi: 10.1007/978-0-387-78747-3.

[12]    K. Pietrak, P. Łapka, and M. Kujawińska, "Inverse analysis for the identification of temporal and spatial characteristics of a short super-Gaussian laser pulse interacting with a solid plate," *Int. J. Therm. Sci.*, vol. 134, pp. 585–593, 2018, doi: 10.1016/j.ijthermalsci.2018.08.040.

[13]    M. N. Ozisik and H. R. B. Orlande, *Inverse heat transfer*. New York: Taylor&Francis, 2000.

[14]    P. Łapka, K. Pietrak, M. Kujawińska, and M. Malesa, "Development and validation of an inverse method for identification of thermal characteristics of a short laser pulse," *Int. J. Therm. Sci.*, vol. 150, p. 106240, 2020, doi: 10.1016/j.ijthermalsci.2019.106240.

[15]    J. V Beck and K. A. Woodbury, "Inverse problems and parameter estimation: integration of measurements and analysis," *Meas. Sci. Technol.*, vol. 9, no. 6, pp. 839–847, 1998, doi: 10.1088/0957-0233/9/6/001.

[16]    J. Taler and D. Taler, "Measurement of Heat Flux and Heat Transfer Coefficient," in *Heat Flux, processes, measurement techniques and applications*, G. Cirimele and M. D'Elia, Eds., New York: Nova Science Publishers, Inc, 2012, pp. 1–104.

[17]    J. Taler and D. Taler, "Surface-heat transfer measurements using transient techniques," in *Encyclopedia of Thermal Stresses*, R. B. Hetnarski, Ed., Dordrecht Heidelberg New York London: Springer, 2014, pp. 4774–4784.

[18]    M. S. Gadala and S. Vakili, "Assessment of Various Methods in Solving Inverse Heat Conduction Problems," in *Heat Conduction - Basic Research*, V. Vikhrenko, Ed., InTech, 2011. doi: 10.5772/28890.

[19]    J. Krejsa, K. A. Woodbury, J. D. Ratliff, and M. Raudensky, "Assessment of strategies and potential for neural networks in the inverse heat conduction problem," *Inverse Probl. Eng.*, vol. 7, no. 3, pp. 197–213, 1999, doi: 10.1080/174159799088027694.

[20]    X. Wang and M. Yao, "Neural networks for solving the inverse heat transfer problem of continuous casting mould," in *2011 Seventh International Conference on Natural Computation*, 2011, pp. 791–794. doi: 10.1109/ICNC.2011.6022280.

[21]    M. Grabarczyk and P. Furmański, "Predicting the effective thermal conductivity of dry granular media using artificial neural networks," *J. Power Technol.*, vol. 93, no. 2, pp. 59–66, 2011, [Online]. Available: http://papers.itc.pw.edu.pl/index.php/JPT/article/view/367

[22] K. Goudarzi, A. Moosaei, and M. Gharaati, "Applying artificial neural networks (ANN) to the estimation of thermal contact conductance in the exhaust valve of internal combustion engine," *Appl. Therm. Eng.*, vol. 87, pp. 688–697, 2015, doi: 10.1016/J.APPLTHERMALENG.2015.05.060.

[23] B. Czél, K. A. Woodbury, and G. Gróf, "Simultaneous estimation of temperature-dependent volumetric heat capacity and thermal conductivity functions via neural networks," *Int. J. Heat Mass Transf.*, vol. 68, pp. 1–13, 2014, doi: 10.1016/J.IJHEATMASSTRANSFER.2013.09.010.

[24] S. Chudzik, W. Minkina, S. Grys, and W. Minkina, "The application of the artificial neural network and hot probe method in thermal parameters determination of heat insulation materials Part 1-thermal model consideration," in *Proceedings of the IEEE International Conference on Industrial Technology*, Institute of Electrical and Electronics Engineers (IEEE), 2009, pp. 1–5. doi: 10.1109/ICIT.2009.4939511.

[25] S. Chudzik, "Measuring system with a dual needle probe for testing the parameters of heat-insulating materials," *Meas. Sci. Technol.*, vol. 22, no. 7, p. 75703, 2011, doi: 10.1088/0957-0233/22/7/075703.

[26] S. Chudzik, "Measurement of thermal diffusivity of insulating material using an artificial neural network," *Meas. Sci. Technol.*, vol. 23, no. 6, 2012, doi: 10.1088/0957-0233/23/6/065602.

[27] S. Chudzik, W. Minkina, and S. Grys, "The application of the artificial neural network and hot probe method in thermal parameters determination of heat insulation materials Part 2 - application of the neural network," Institute of Electrical and Electronics Engineers (IEEE), 2009, pp. 1–5. doi: 10.1109/icit.2009.4939512.

[28] A. Mirsepahi, L. Chen, and B. O'Neill, "An Artificial Intelligence Solution for Heat Flux Estimation Using Temperature History; A Two-Input/Two-Output Problem," *Chem. Eng. Commun.*, vol. 204, no. 3, pp. 289–294, Mar. 2017, doi: 10.1080/00986445.2016.1253008.

[29] A. Mirsephai, M. Mohammadzaheri, L. Chen, and B. O'Neill, "An artificial intelligence approach to inverse heat transfer modeling of an irradiative dryer," *Int. Commun. Heat Mass Transf.*, vol. 39, no. 1, pp. 40–45, 2012, doi: 10.1016/J.ICHEATMASSTRANSFER.2011.09.015.

[30] A. Mirsepahi, L. Chen, and B. O'Neill, "A comparative artificial intelligence approach to inverse heat transfer modeling of an irradiative dryer," *Int. Commun. Heat Mass Transf.*, vol. 41, pp. 19–27, 2013, doi: 10.1016/J.ICHEATMASSTRANSFER.2012.09.011.

[31] A. Mirsepahi, L. Chen, and B. O'Neill, "A comparative approach of inverse modelling applied to an irradiative batch dryer employing several artificial neural networks," *Int. Commun. Heat Mass Transf.*, vol. 53, pp. 164–173, 2014, doi: 10.1016/J.ICHEATMASSTRANSFER.2014.02.028.

[32] R.-M. Ricardo, H.-L. Juan Manuel, D.-G. Héctor Martín, and P.-V. Arturo, "Use of Artificial Neural Networks for Prediction of Convective Heat Transfer in Evaporative Units," *Ing. Investig. y Tecnol.*, vol. 15, no. 1, pp. 93–101, 2014, doi: 10.1016/S1405-7743(15)30009-3.

[33] R. Romero-Méndez, P. Lara-Vázquez, F. Oviedo-Tolentino, H. M. Durán-García, F. G. Pérez-Gutiérrez, and A. Pacheco-Vega, "Use of Artificial Neural Networks for Prediction of the Convective Heat Transfer Coefficient in Evaporative Mini-Tubes," *Ing. Investig. y Tecnol.*, vol. 17, no. 1, pp. 23–34, 2016, doi: 10.1016/J.RIIT.2016.01.003.

[34] M. Mohanraj, S. Jayaraj, and C. Muraleedharan, "Applications of artificial neural networks for thermal analysis of heat exchangers - A review," *Int. J. Therm. Sci.*, vol. 90, pp. 150–172, 2015, doi: 10.1016/j.ijthermalsci.2014.11.030.

[35] O. Cortés, G. Urquiza, and J. A. Hernández, "Optimization of operating conditions for compressor performance by means of neural network inverse," *Appl. Energy*, vol. 86, no. 11, pp. 2487–2493, 2009, doi: 10.1016/J.APENERGY.2009.03.001.

[36] J. A. Hernández, D. Colorado, O. Cortés-Aburto, Y. El Hamzaoui, V. Velazquez, and B. Alonso, "Inverse neural network for optimal performance in polygeneration systems," *Appl. Therm. Eng.*, vol. 50, no. 2, pp. 1399–1406, 2013, doi: 10.1016/j.applthermaleng.2011.12.041.

[37] Y. El Hamzaoui, B. Ali, J. A. Hernandez, O. Cortés-Aburto, and O. Oubram, "Search for Optimum Operating Conditions for a Water Purification Process Integrated to a Heat Transformer with Energy Recycling using Artificial Neural Network Inverse Solved by Genetic and Particle Swarm Algorithms," *Chem. Prod. Process Model.*, vol. 7, no. 1, 2012, doi: 10.1515/1934-2659.1614.

[38]    S. Samarasinghe, *Neural Networks for Applied Sciences and Engineering: From Fundamentals to Complex Pattern Recognition*. Boca Raton: Auerbach, 2007.

[39]    Engineering ToolBox, "Metals - Melting Temperatures." 2005. Accessed: May 28, 2018. [Online]. Available: https://www.engineeringtoolbox.com/melting-temperature-metals-d_860.html

[40]    Y. A. Cengel and A. J. Ghajar, *Heat and Mass Transfer, Fundamentals & Applications*, 5th-th ed. New York: McGraw-Hill Education, 2015.

[41]    "FLIR SC7500 instrument specifications," *Flir Systems Inc.* 2014. [Online]. Available: https://issuu.com/netservis/docs/flir-sc7500

[42]    A. Gulli and S. Pal, *Deep Learning with Keras*. Packt Publishing Ltd, 2017.

[43]    S. Ruder, "An overview of gradient descent optimization algorithms," 2016, [Online]. Available: http://arxiv.org/abs/1609.04747

[44]    B. Ding, H. Qian, and J. Zhou, "Activation functions and their characteristics in deep neural networks," in *Proceedings of the 30th Chinese Control and Decision Conference, CCDC 2018*, Institute of Electrical and Electronics Engineers Inc., 2018, pp. 1836–1841. doi: 10.1109/CCDC.2018.8407425.

[45]    C. Nwankpa, W. Ijomah, A. Gachagan, and S. Marshall, "Activation Functions: Comparison of trends in Practice and Research for Deep Learning," 2018, [Online]. Available: http://arxiv.org/abs/1811.03378

[46]    D.-A. Clevert, T. Unterthiner, and S. Hochreiter, "Fast and Accurate Deep Network Learning by Exponential Linear Units (ELUs)," *4th Int. Conf. Learn. Represent. ICLR 2016 - Conf. Track Proc.*, 2015, [Online]. Available: http://arxiv.org/abs/1511.07289

[47]    G. Klambauer, T. Unterthiner, A. Mayr, and S. Hochreiter, "Self-Normalizing Neural Networks," *Adv. Neural Inf. Process. Syst.*, vol. 2017-Decem, pp. 972–981, 2017, [Online]. Available: http://arxiv.org/abs/1706.02515

[48]    Keras Documentation, "Earlystopping." 2020. Accessed: Jul. 29, 2020. [Online]. Available: https://keras.io/api/callbacks/early_stopping/

[49]    R. J. Hyndman and A. B. Koehler, "Another look at measures of forecast accuracy," *Int. J. Forecast.*, vol. 22, no. 4, pp. 679–688, 2006, doi: 10.1016/J.IJFORECAST.2006.03.001.

[50]    J. Brownlee, "How to Configure the Number of Layers and Nodes in a Neural Network." 2018. Accessed: Jul. 30, 2020. [Online]. Available: https://machinelearningmastery.com/how-to-configure-the-number-of-layers-and-nodes-in-a-neural-network/

[51]    N. Srivastava, G. Hinton, A. Krizhevsky, and R. Salakhutdinov, "Dropout: A Simple Way to Prevent Neural Networks from Overfitting," 2014. doi: 10.5555/2627435.2670313.

[52]    D. Stathakis, "How many hidden layers and nodes?," *Int. J. Remote Sens.*, vol. 30, no. 8, pp. 2133–2147, 2009, doi: 10.1080/01431160802549278.

[53]    V. Kůrková, "Kolmogorov's theorem and multilayer neural networks," *Neural Networks*, vol. 5, no. 3, pp. 501–506, 1992, doi: 10.1016/0893-6080(92)90012-8.

[54]    G. Cybenko, "Approximation by superpositions of a sigmoidal function," *Math. Control. Signals, Syst.*, vol. 2, no. 4, pp. 303–314, 1989, doi: 10.1007/BF02551274.

[55]    W. S. Sarle, "Should I normalize/standardize/rescale the data?" 2002. Accessed: Aug. 24, 2020. [Online]. Available: http://www.faqs.org/faqs/ai-faq/neural-nets/part2/section-16.html

[56]    S. Ioffe and C. Szegedy, "Batch normalization: Accelerating deep network training by reducing internal covariate shift," in *32nd International Conference on Machine Learning, ICML 2015*, International Machine Learning Society (IMLS), 2015, pp. 448–456. [Online]. Available: https://arxiv.org/abs/1502.03167v3

[57]    J. Brownlee, "How to use Data Scaling Improve Deep Learning Model Stability and Performance." 2019. Accessed: Jul. 29, 2020. [Online]. Available: https://machinelearningmastery.com/how-to-improve-neural-network-stability-and-modeling-performance-with-data-scaling/

[58] S. Nayak, B. B. Misra, and H. S. Behera, "Impact of Data Normalization on Stock Index Forecasting," *Int. J. Comput. Inf. Syst. Ind. Manag. Appl.*, vol. 2014, pp. 257–269, 2014.

[59] Tibo Spotfire Documentation, "Normalization by subtracting the median." 2014. Accessed: Aug. 25, 2020. [Online]. Available: https://docs.tibco.com/pub/spotfire/6.5.0/doc/html/norm/norm_subtract_the_median.htm

[60] J. Jones, "Stats: Measures of Central Tendency." 2020. Accessed: Aug. 25, 2020. [Online]. Available: https://people.richland.edu/james/lecture/m170/ch03-ave.html

[61] K. Pietrak, P. Łapka, and M. Kujawińska, "Numerical tests of an inverse method for the characterization of pulsed and continuous laser beams," in *Contemporary Issues of Heat and Mass Transfer*, M. Kruzel and W. Kuczyński, Eds., Koszalin University of Technology, 2019, pp. 481–502.